\documentclass[%
 reprint,
superscriptaddress,
 amsmath,amssymb,
 aps,floatfix,
pra,
]{revtex4-2}

\usepackage{graphicx}% Include figure files
\usepackage{dcolumn}% Align table columns on decimal point
\usepackage{bm}% bold math
\usepackage{cancel}
\usepackage{hyperref}% add hypertext capabilities
\usepackage{graphicx}
\usepackage{subfigure}
\usepackage{amsmath}
\usepackage{multirow}
\usepackage{float}
\usepackage{soul} %added by LQ for certain fonts
\usepackage{xcolor} %added by LQ for certain text color

\begin{document}

\preprint{APS/123-QED}

\title{
Entanglement distillation based on polarization and frequency hyperentanglement
}

\author{Dan Xu}
\email{xudan.xu@utoronto.ca}
 \affiliation{Department of Electrical and Computer Engineering, University of Toronto, Toronto, ON, M5S 3G4 Canada}

\author{Changjia Chen}
 \affiliation{Department of Electrical and Computer Engineering, University of Toronto, Toronto, ON, M5S 3G4 Canada}

\author{Brian T. Kirby}
 \affiliation{DEVCOM US Army Research Laboratory, Adelphi, MD 20783, USA}
 \affiliation{Tulane University, New Orleans, LA 70118, USA}
 
\author{Li Qian}
\email{l.qian@utoronto.ca}
 \affiliation{Department of Electrical and Computer Engineering, University of Toronto, Toronto, ON, M5S 3G4 Canada}

\date{\today}
             
\begin{abstract}
Entanglement distillation has many applications in quantum information processing and is an important tool for improving the quality and efficiency of quantum communication, cryptography, computing, and simulation. We propose an entanglement distillation scheme using only one pair of polarization-frequency hyperentangled photons, which can be equivalently viewed as containing two pairs of entangled logical qubits: a pair of polarization-entangled qubits and a pair of frequency-entangled qubits. To perform the required CNOT operation between the two qubits we consider the use of a polarization-dependent frequency converter. Compared to past methods of entanglement distillation that relied on polarization and spatial-mode/energy-time degree of freedom, the utilization of frequency-encoded qubits offers an advantage in that it is immune to bit-flip errors when the channel is linear. After distillation, the fidelity of polarization entanglement can be significantly improved by sacrificing the frequency degree of freedom. Through simulation, we show that high fidelity gains, large yield, and high distillation rate can be achieved. Our distillation scheme is simple to implement with current technologies, compatible with existing telecommunication fiber networks, and is a promising approach for achieving efficient quantum communication.
\end{abstract}
%\showkeys{Quantum entanglement, hyperentanglement, entanglement distillation, fidelity, quantum communication.}%Use showkeys class option if keyword

\maketitle

\section{\label{intro}Introduction}
% Quantum communication/quantum network
% Need for entanglement distribution over large distances
Distribution of entangled states between distant locations will be essential for the future large-scale realization of quantum communication schemes such as quantum teleportation\,\cite{pirandola_advances_2020}, quantum dense coding\,\cite{guo_advances_2019}, quantum cryptography\,\cite{portmann_security_2022}, quantum digital signature\,\cite{zhang_proof--principle_2018,pelet_unconditionally_2022}, coin-flipping\,\cite{neves_experimental_2022} as well as quantum-enhanced metrological schemes\,\cite{pezze_quantum_2018} and quantum computation\,\cite{zhong_quantum_2020}. However, noise in the quantum channels and interactions with the environment are unavoidable, leading to decoherence and the degradation of entanglement. 

% Review of ED key papers: Benett96, Pan2001,...
% low efficiency of two-copy ED.
To counteract such detrimental effects, entanglement distillation (ED), also referred to as entanglement purification, was introduced. The ED protocol was first proposed by Bennett et al. in 1996 and was further developed in both theory and experiment\,\cite{bennett_purification_1996,pan_entanglement_2001,pan_experimental_2003,yamamoto_experimental_2003,walther_quantum_2005,chen_experimental_2017}. In the original (or two-copy) ED protocol, two imperfectly entangled pairs are consumed to generate, with some nonzero probability, a single entangled pair of a higher fidelity to a maximally entangled state. To perform ED, two different and independent entangled pairs are shared between remote users simultaneously. Once shared, the remote users conduct distillation by each performing a local CNOT operation between their portion of the two entangled states accompanied by the detection of one of the pairs which herald success or failure. The heralded results based on the detection of the target pairs will lead to a higher probability of retaining the desirable entangled states and discarding the noise states. Such distillation processes can be repeated to obtain high-fidelity entangled states by sacrificing a large amount of low-fidelity pairs. Experimental implementations of such ED protocols, however, face the challenge of low efficiency and low distillation rates, as they are substantially limited by the low transmission probabilities of multiple photon pairs. 
%\textcolor{blue}{After two remote parties perform a controlled-not (CNOT) (or equivalent) operation between two pairs of photons, a source and a target, they measure the target pair to decide whether to discard the source pair or not.} 

% Efficiency can be dramatically improved by using single-copy ED, i.e. using hyperentanglement
More recently, to overcome low efficiency, new ED protocols were proposed using only one copy of imperfect hyperentangled states\,\cite{hu_long-distance_2021,ecker_experimental_2021,sheng_one-step_2010,yan_advances_2023}. Hyperentanglement, the simultaneous and independent entanglement of quantum particles in multiple degrees of freedom (e.g. frequency, time-bin, polarization, orbital angular momentum), significantly expands the dimensionality of the Hilbert space of biphotons\,\cite{barreiro_generation_2005,barbieri_polarization-momentum_2005,erhard_advances_2020}. In practice, hyperentanglement can be naturally produced in certain spontaneous parametric downconversion (SPDC) processes\,\cite{chen_compensation-free_2017,chen_recovering_2020,chen_telecom-band_2022}. By constructing a ``CNOT" gate between different degrees of freedom (DoFs), e.g., between polarization and spatial-mode or between polarization and energy-time DoFs, the entanglement in the latter DoF can be sacrificed to distill the polarization entanglement. Such hyperentanglement-based ED uses only one pair of photons to distill out a higher quality of entanglement in one DoF by sacrificing the other DoF, resulting in a dramatically improved efficiency. Though hyperentanglement-based ED is efficient, the process cannot be repeated indefinitely since, after one purification step, the entanglement in the other DoF is sacrificed. Therefore, it is desirable to obtain a significant improvement in fidelity in one step.  
% Review of ED with hyperentanglement: pros and cons
% Motivation: we have a PPSF-based polarization-frequency hyperentangled photon source in the lab, which has good performance.
The ED protocol in Ref.\,\cite{ecker_experimental_2021} using the energy-time DoF is more robust in in-field distribution than the spatial mode DoF used in Ref.\,\cite{hu_long-distance_2021}, because the former does not need to transmit multiple spatial modes coherently over a long distance. On the other hand, the measurement of energy-time DoF requires modified Franson interferometers, which need to be stabilized at both remote sites, increasing the complexity of the distillation setup signficantly. 

\begin{figure}[!t]
\centering
\includegraphics[width=3.4in]{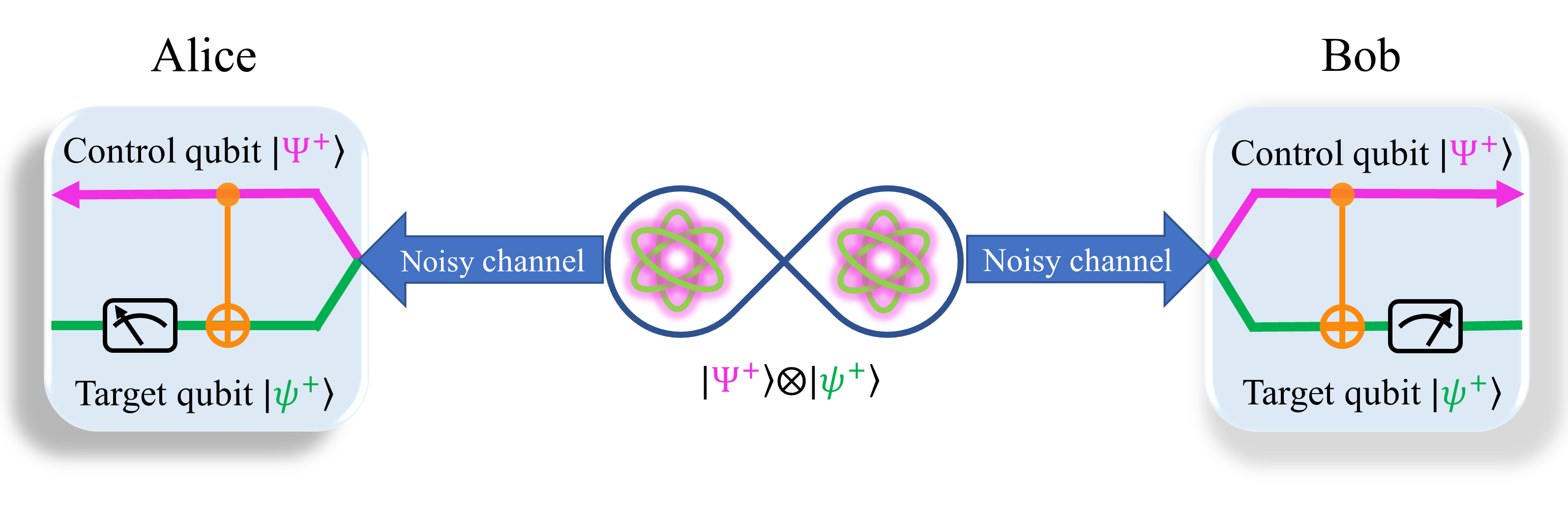}
\caption{A conceptual illustration of entanglement distillation protocol using polarization-frequency hyperentanglement. The hyperentangled state $|\Psi^+\rangle\otimes|\psi^+\rangle$ is first generated by an entanglement source and then distributed to Alice and Bob over two quantum channels. Alice and Bob apply a ``CNOT" gate between two entangled subspaces encoded in the polarization (control qubit) and frequency (target qubit) degrees of freedom of a single photon pair. The successful distillation of the polarization entanglement is achieved after postselection based on the measurement results in the frequency DoF.}
\label{fig1}
\end{figure}

% This paper proposes an ED protocol...
Here we propose an ED protocol exploiting hyperentanglement in the polarization and frequency DoFs, which is both robust to channel degradation and eliminates the need for interferometric stability. Entanglement distillation relies on the ``CNOT" gates with the polarization DoF as the control qubit and the frequency DoF as the target qubit. A conceptual diagram is shown in Fig.\,\ref{fig1}. We analyze our distillation scheme regarding fidelity, yield, and distillation rate. We compare it with the ED protocols using two copies\,\cite{bennett_purification_1996}, as well as using the polarization-energy-time (PET)\,\cite{ecker_experimental_2021} and the polarization-spatial mode (PSM)\,\cite{hu_long-distance_2021} hyperentanglement. Our entanglement distillation approach is highly efficient, robust, and easy to integrate into existing telecommunication fiber networks and can thus become a vital building block of future quantum internet.

\section{proposal}
% Conceptual diagram
Our proposed polarization entanglement distillation protocol uses only one pair of hyperentanglement in the polarization and frequency DoFs. We consider control and target implemented as polarization and frequency qubits, respectively. We denote the state of a horizontally polarized photon by $|H\rangle$ and the state of a vertically polarized photon by $|V\rangle$. As shown in Fig.\,\ref{fig1}, the hyperentangled photon source generates one pair of hyperentangled states which can be written as
\begin{equation}
|\phi\rangle=|\Psi^+\rangle_{ab}\otimes|\psi^+\rangle_{ab}.
\label{eq1}
\end{equation}
$|\Psi^+\rangle_{ab}$ is the polarization Bell state
\begin{equation}
|\Psi^+\rangle_{ab}=\frac{1}{\sqrt{2}}(|H\rangle_{a}|V\rangle_{b}+|V\rangle_{a}|H\rangle_{b}),
\label{eq2}
\end{equation}
and $|\psi^+\rangle_{ab}$ is the frequency Bell state 
\begin{equation}
|\psi^+\rangle_{ab}=\frac{1}{\sqrt{2}}(|\omega_s\rangle_a|\omega_i\rangle_b+|\omega_i\rangle_a|\omega_s\rangle_b)
\label{eq3}
\end{equation}
where $\omega_s,\omega_i$ are the frequencies of the signal and idler photons, and the photons at Alice’s and Bob’s locations are denoted by $a$ and $b$, respectively. The hyperentangled photons are distributed to Alice and Bob across a quantum channel. During distribution over a noisy channel, a phase flip (PF) error ($|\Psi^-\rangle_{ab}$) and a bit flip (BF) error ($|\Phi^+\rangle_{ab}$), as well as a bit-phase flip (BPF) error ($|\Phi^-\rangle_{ab}$), may be introduced to the polarization state. Due to the energy conservation of the SPDC process, the frequency Bell states could only be in $|\psi^+\rangle$ and $|\psi^-\rangle$. Additionally, assuming the channel is linear, no BF ($|\phi^+\rangle_{ab}$) or BPF error ($|\phi^-\rangle_{ab}$) can occur in frequency qubits. The assumption of a linear channel is reasonable if the channel is used to transmit quantum signals only, which are many orders of magnitude too weak to induce a nonlinear phase or frequency change. As we shall see later, the robustness in transmitting frequency qubits results in a drastic improvement in fidelity gain for the polarization qubits after distillation. For this reason, it is desirable to use frequency as an auxiliary DoF.  

Following transmission over a noisy channel, the initial hyperentangled state becomes a mixed state as
\begin{equation}
\rho_{ab}=\rho_{ab}^{p}\otimes\rho_{ab}^{f},
\label{eq4}
\end{equation}
with
\begin{equation}
\begin{split}
\rho_{ab}^{p} & = F_p|\Psi^+\rangle_{ab}\langle\Psi^+|+A|\Psi^-\rangle_{ab}\langle\Psi^-| \\
&{\hspace{12pt}} +B|\Phi^+\rangle_{ab}\langle\Phi^+|+C|\Phi^-\rangle_{ab}\langle\Phi^-|
\end{split}
\label{eq5}
\end{equation}
and
\begin{equation}
\rho_{ab}^{f}=F_f|\psi^+\rangle_{ab}\langle\psi^+|+(1-F_f)|\psi^-\rangle_{ab}\langle\psi^-|,
\label{eq6}
\end{equation}
where $F_p$, $A$, $B$, $C$ are the probabilities of polarization states $|\Psi^+\rangle$, $|\Psi^-\rangle$, $|\Phi^+\rangle$, $|\Phi^-\rangle$, respectively, and $F_p+A+B+C=1$. $F_f$, $(1-F_f)$ are the probabilities of frequency states $|\psi^+\rangle$ and $|\psi^-\rangle$, respectively. Table\,\ref{table_fidelity} lists all the error types in the polarization and frequency states, as well as their corresponding probabilities.

\begin{table*}[!t]
\renewcommand{\arraystretch}{1.6}
\caption{All the error types in the polarization and the frequency and their corresponding probabilities before and after distillation. The postselection is based on the frequency state: the photon pairs in the state $|\phi^{\pm}\rangle_{AB}$ will be kept and in the state $|\psi^{\pm}\rangle_{AB}$ will be discarded.}
\label{table_fidelity}
\centering
\begin{tabular}{cccc|ccc|c}
\hline
\multicolumn{4}{c|}{Initial state} & \multicolumn{3}{c|}{After ``CNOT"} &  Postselection\\
\hline
Polarization & Probability & Frequency & Probability & Polarization & Frequency & Probability & Decision\\
\hline
\multirow{2}{2em}{$|\Psi^+\rangle$} & \multirow{2}{2em}{$F_p$} & $|\psi^+\rangle$ & $F_f$ & $|\Psi^+\rangle$ & $|\phi^+\rangle$ & $F_pF_f$ & keep\\
 & & $|\psi^-\rangle$ & $1-F_f$ & $|\Psi^-\rangle$ & $|\phi^-\rangle$ & $F_p(1-F_f)$ & keep\\
\hline
$|\Psi^-\rangle$ & \multirow{2}{2em}{$A$} & $|\psi^+\rangle$ & $F_f$ & $|\Psi^-\rangle$ & $|\phi^+\rangle$ & $AF_f$ & keep\\
(PF error) & & $|\psi^-\rangle$ & $1-F_f$ & $|\Psi^+\rangle$ & $|\phi^-\rangle$ & $A(1-F_f)$ & keep\\
\hline
$|\Phi^+\rangle$ & \multirow{2}{2em}{$B$} & $|\psi^+\rangle$ & $F_f$ & $|\Phi^+\rangle$ & $|\psi^+\rangle$ & $BF_f$ & discard\\
(BF error) & & $|\psi^-\rangle$ & $1-F_f$ & $|\Phi^-\rangle$ & $|\psi^-\rangle$ & $B(1-F_f)$ & discard\\
\hline
$|\Phi^-\rangle$ & \multirow{2}{2em}{$C$} & $|\psi^+\rangle$ & $F_f$ & $|\Phi^-\rangle$ & $|\psi^+\rangle$ & $CF_f$ & discard\\
(BPF error) & & $|\psi^-\rangle$ & $1-F_f$ & $|\Phi^+\rangle$ & $|\psi^-\rangle$ & $C(1-F_f)$ & discard\\
\hline
\end{tabular}
\end{table*}

% description of the protocol
Our distillation protocol is based on a polarization-dependent frequency converter which is essentially a single-photon ``CNOT" gate between the polarization and frequency DoFs. We start the explanation of our purification scheme by discussing an ideal, simplified example. Suppose that the two ``CNOT" gates (frequency converters) utilized in this protocol on Alice's and Bob's sides are identical and their frequency conversion efficiency $\eta=100\%$. The distillation protocol is composed of the following steps:

\textbf{Step 1:} Alice and Bob obtain their entangled photons from a prime source in a hyperentangled state described by Eq.\,\eqref{eq1}. 

\textbf{Step 2:} Alice and Bob perform a ``CNOT" operation on their received photons. Depending on the polarization state (control qubit) of the photon, the frequency (target qubit) of the photon is either left unchanged or converted. The logic table of the ``CNOT" operation is
\begin{equation}
\begin{split}
|\omega_s\rangle|H\rangle & \Rightarrow|\omega_s\rangle|H\rangle,\\
|\omega_i\rangle|H\rangle & \Rightarrow|\omega_i\rangle|H\rangle,\\ 
|\omega_s\rangle|V\rangle & \Rightarrow|\omega_i\rangle|V\rangle,\\
|\omega_i\rangle|V\rangle & \Rightarrow|\omega_s\rangle|V\rangle. 
\end{split}
\label{eq7}
\end{equation}
i.e., the frequency mode is flipped only if the photon incident on the frequency converter is vertically polarized. In practice, this can be achieved through a polarization-dependent nonlinear frequency conversion process involving a strong classical pump. Temporal distinguishability introduced between the H and V polarizations due to the birefringence of the nonlinear waveguide can be later erased in a compensation step.
%\textcolor{red}{[Maybe worth including a citation for previous uses of frequency-conversion as a CNOT?]}

\textbf{Step 3:} After the ``CNOT" operation, it becomes a new mixed state as listed in Table\,\ref{table_fidelity}. The frequency state will be flipped ($|\psi^{\pm}\rangle\Rightarrow|\phi^{\pm}\rangle$) when paired with the $|\Psi^{\pm}\rangle$ states in polarization DoF. 

\textbf{Step 4:} Alice and Bob's postselection is based on the frequency qubit after performing the CNOT operation. The cases in which the photon pairs are in the frequency state $|\psi^{\pm}\rangle_{ab}$ are discarded in postselection. It is worth noting that distinguishing between $|\psi^+\rangle_{ab}$ and $|\psi^-\rangle_{ab}$ states during measurement is unnecessary because both will be discarded. The cases in which the photon pairs are in the frequency state $|\phi^{\pm}\rangle_{ab}$ are kept in postselection. As shown in Table\,\ref{table_fidelity}, all BF and BPF errors in polarization can be identified and discarded. 

By applying our distillation procedure, they can thus create a new ensemble
\begin{equation}
\rho_{ab}^{\prime p} = F_p^{\prime}|\Psi^+\rangle_{ab}\langle\Psi^+|+(1-F_p^{\prime})|\Psi^-\rangle_{ab}\langle\Psi^-|
\label{eq8}
\end{equation}
with a larger fraction $F_p^{\prime}>F_p$ of entangled pairs in the desired polarization state $|\Psi^+\rangle_{ab}$ than before the distillation. We now analyze them in more detail to show that our distillation scheme also works for various noise scenarios.

\section{Analysis}
The success of the distillation protocol can be evaluated by comparing the noisy state $\rho_{ab}^{p}$ and the distilled state $\rho_{ab}^{\prime p}$ with each other. For this purpose, we utilize the quantum state fidelity $F$ to the $|\Psi^+\rangle_{ab}$ state before and after the distillation. Here we only consider phase-flip error ($\rho^f_\text{err}=(1-F_f)|\psi^-\rangle_{ab}\langle\psi^-|$) to the frequency DoF and its state can be described by Eq.\,\eqref{eq6} with fidelity $F_f$. It does not reduce generality, as this is the only error possible in a linear channel. In regards to the polarization DoF, to comprehensively evaluate the efficiency of the protocol, we take into account all potential errors.

\begin{figure*}[!t]
\centering
\includegraphics[width=0.8\linewidth]{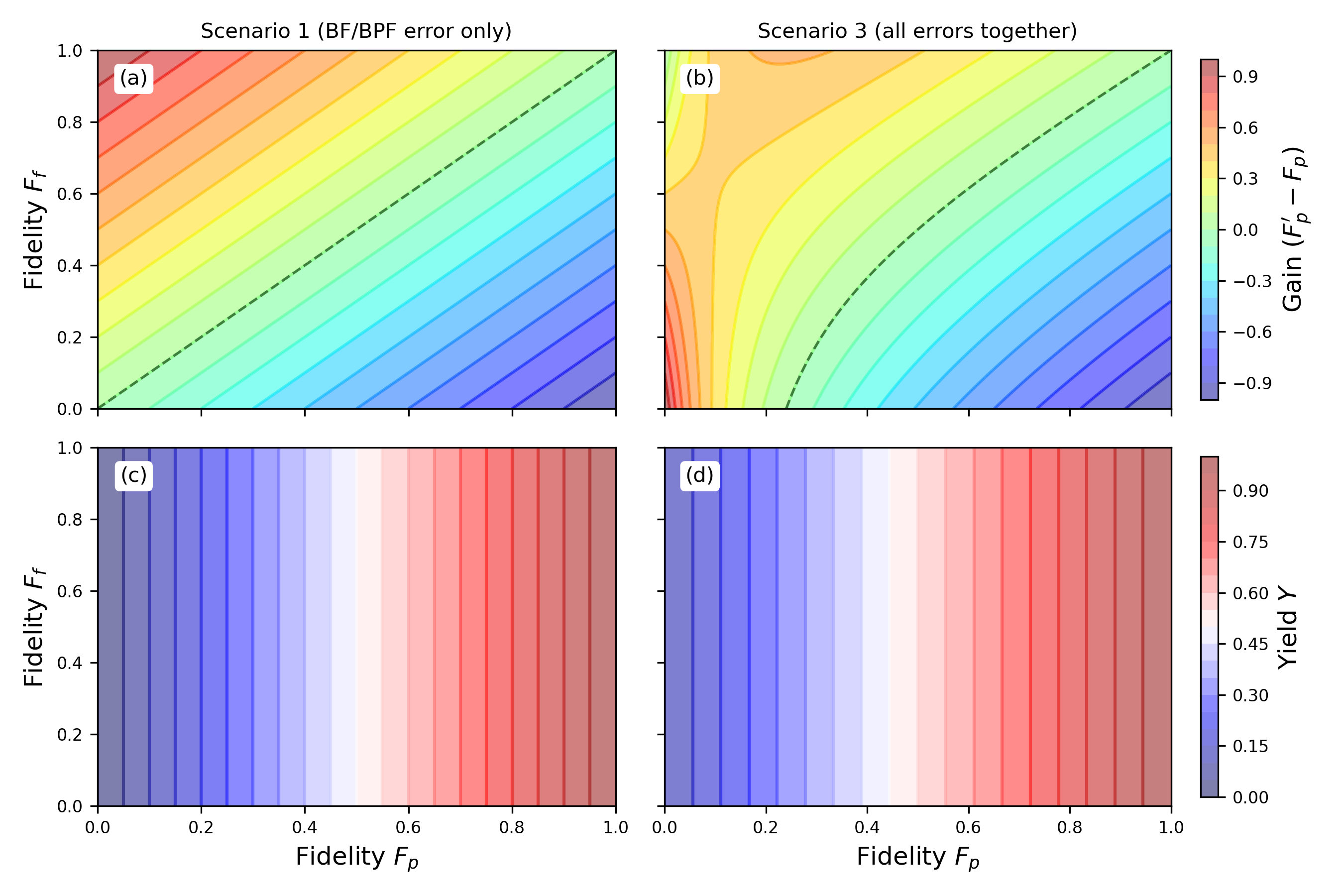}
\caption{The distilled polarization fidelity gain $G$ and yield $Y$ for (a,c) \textbf{Scenario 1} (BF or BPF error only) and (b,d) \textbf{Scenario 3} (all errors together) ($A=10\%$), respectively. Here we only consider PF error in frequency DoF because the channel is linear and BF \& BPF errors for the frequency qubit cannot happen. Nevertheless, given that the main focus is on polarization while considering the other DoF as an auxiliary, we define the gain as $G=F^{\prime}_p-F_p$. The dashed line separates the region of distillable states ($G > 0$) from the region of nondistillable states ($G < 0$). In these scenarios, the Yield $Y$ is independent of the fidelity of frequency qubit $F_f$.}
\label{Fig_PF}
\end{figure*}

\textbf{Scenario 1 (BF or BPF error only)}: we first introduce a BF error ($\rho^p_\text{err}=B|\Phi^+\rangle_{ab}\langle\Phi^+|$) to the polarization DoF, i.e., $B=1-F_p, A=C=0$. After the ``CNOT" operation and postselection, the polarization fidelity of the distilled state is $F_p^{\prime}=F_f$. We also examine the BPF error ($\rho^p_\text{err}=C|\Phi^-\rangle_{ab}\langle\Phi^-|$) as an additional error type, i.e., $C=1-F_p, A=B=0$. We observe that this error shows the same fidelity characteristics $F_p^{\prime}=F_f$ as the BF error. Thus, in the case of only BF or only BPF error, or both ($\rho^p_\text{err}=B|\Phi^+\rangle_{ab}\langle\Phi^+| + C|\Phi^-\rangle_{ab}\langle\Phi^-|$, i.e., $A=0$), the polarization fidelity of the distilled state will be dramatically improved. Theoretically, our distillation scheme can distill the polarization back to the ideal Bell state $\rho^{\prime p} = |\Psi^+\rangle_{ab}\langle\Psi^+|$ if $F_f^{\prime}=1$. 

\textbf{Scenario 2 (PF error only)}: In this scenario, we consider all errors in the polarization DoF are PF errors, i.e., $\rho^p_\text{err}=A|\Psi^-\rangle_{ab}\langle\Psi^-|$, and $A=1-F_p, B=C=0$. After the ``CNOT" operation and postselection, the polarization fidelity of the distilled state is $F_p^{\prime}=F_pF_f+(1-F_p)(1-F_f)$. As seen from Table\,\ref{table_fidelity}, the PF error does not result in any discarded states, and therefore the distillation protocol as listed in Table\,\ref{table_fidelity} is not effective when $F_p>0.5$. However, for this scenario, we can modify the protocol as follows: Apply a Hadamard gate in the polarization DoF, such that $|\Phi^+\rangle\leftrightarrow|\Phi^+\rangle$, $|\Phi^-\rangle\leftrightarrow|\Psi^+\rangle$, $|\Psi^-\rangle\leftrightarrow|\Psi^-\rangle$. We then apply the CNOT gate, and keep all $|\psi^\pm\rangle$ states while discarding all $|\phi^\pm\rangle$ states. This will yield the same fidelity as in Scenario 1, i.e., $F_p^{\prime}=F_f$. 

\textbf{Scenario 3 (all errors together)}: We consider that all types of errors coexisting in the polarization DoF, i.e., $\rho^p_\text{err}=A|\Psi^-\rangle_{ab}\langle\Psi^-|+B|\Phi^+\rangle_{ab}\langle\Phi^+|+C|\Phi^-\rangle_{ab}\langle\Phi^-|$. Applying the same protocol as listed in Table\,\ref{table_fidelity}, the polarization fidelity after distillation is $F_p^{\prime}=\frac{F_pF_f+A(1-F_f)}{F_p+A}$. This protocol works well if the BF or BPF error predominates (i.e., $A$ is small). When considering a range of $A$ values (see Appendix), the optimal protocol depends on the trade-off between the fidelity gain and yield. Specifically, when PF error dominates, it is more advantageous to apply the modified protocol mentioned in Scenario 2. 

% Gain
The aforementioned theoretical analysis anticipates the adaptability of our distillation protocol to a multitude of noise scenarios. Since polarization qubit is the target while considering frequency qubit as an auxiliary, we use the fidelity gain $G=F^{\prime}_p-F_p$ as our figure of merit. It is evident from the results depicted in Fig.\,\ref{Fig_PF}(a) that, for \textbf{Scenario 1}, the fidelity gain $G$ remains positive across the entire range of $F_p\in[0,1]$, specifically when $F_f>F_p$. If the frequency state maintains a high fidelity of $F_f\approx1$ throughout the linear channel, the estimated theoretical fidelity gain after distillation amounts to $G=0.5$ for an initially mixed state with $F_p=0.5$. It significantly surpasses the fidelity of the polarization state before distillation. For \textbf{Scenario 3}, Fig.\,\ref{Fig_PF}(b) reveals analogous outcomes to those observed in \textbf{Scenario 1}, particularly when considering a small value of $A$ at 10\%. 

\begin{figure*}[!t]
\centering
\includegraphics[width=0.8\linewidth]{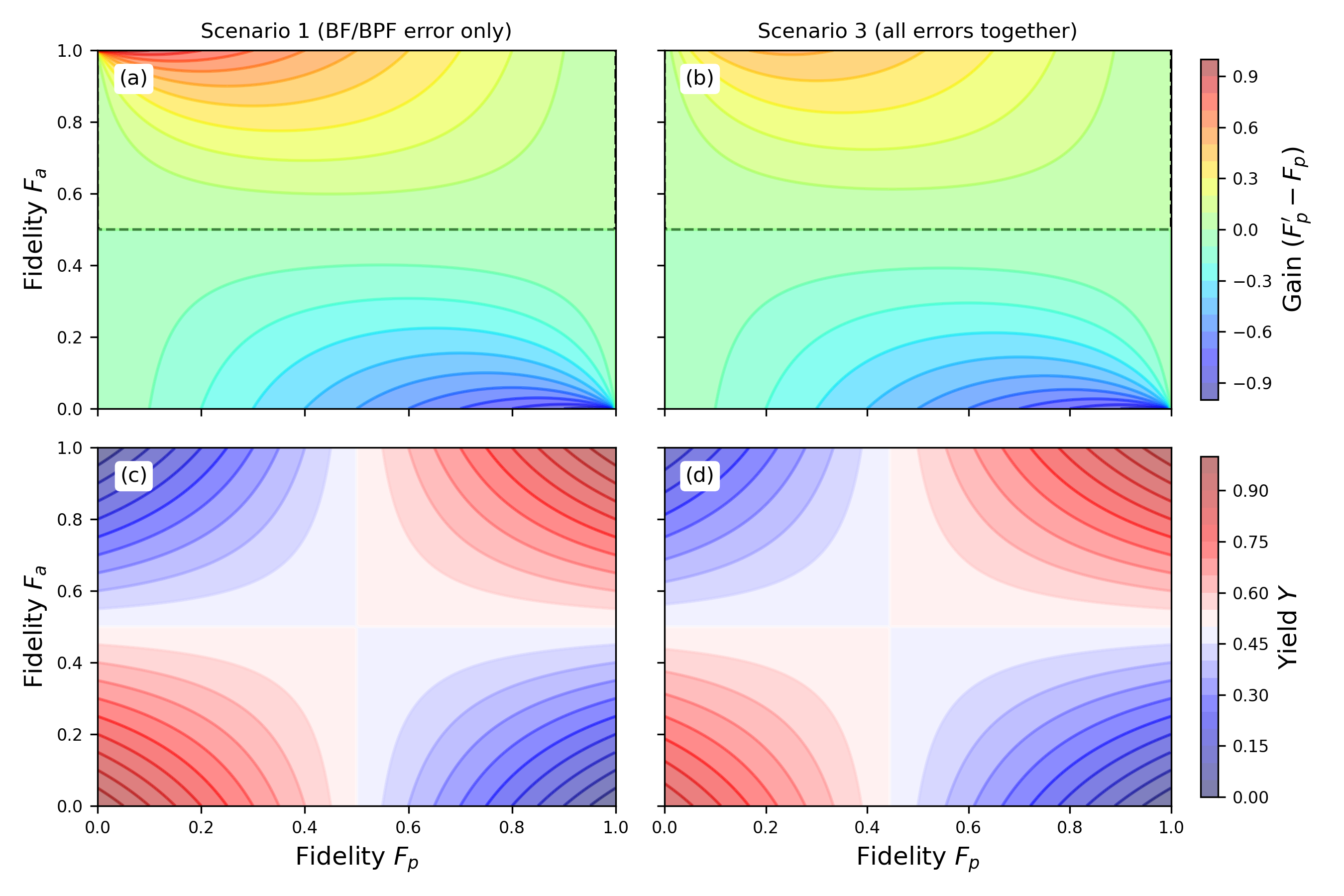}
\caption{The distilled polarization fidelity gain $G$ and yield $Y$ for (a,c) \textbf{Scenario 1} (BF or BPF error only) and (b,d) \textbf{Scenario 3} (all errors together) ($A=10\%$) when considering BF/BPF error in the auxiliary DoF, e.g. spatial-mode\,\cite{hu_long-distance_2021} or energy-time\,\cite{ecker_experimental_2021} or frequency (this work). For the three hyperentanglement-based protocols, they obtain the same results and a positive gain $G>0$ only when the auxiliary qubit is in the range $F_a\in[0.5,1]$.}
\label{Fig_BF}
\end{figure*}

% Yield
Another important figure of merit for distillation protocols is the yield, $Y$, which quantifies the ratio of successfully distilled photon pairs to the total count of coincident photon pairs. In the context of \textbf{Scenario 1}, the yield $Y$ exhibits a linear relationship with the initial fidelity of the polarization state, $Y=F_p$, while remaining unaffected by the fidelity of the frequency state $F_f$, as illustrated in Fig.\,\ref{Fig_PF}(c). In the case of \textbf{Scenario 3}, the yield $Y=F_p+A$ demonstrates analogous trends when $A$ takes on a small value. 

The distillation protocols utilizing PET and PSM hyperentanglement take into account BF error within the polarization state and BF/BPF errors in the other DoF\,\cite{hu_long-distance_2021,ecker_experimental_2021}. For the sake of consistency, we also examine scenarios wherein the frequency state $|\psi^+\rangle$ with an admixed BF contribution $|\phi^+\rangle$. In such instances, these three hyperentanglement-based distillation protocols yield identical fidelity gain and yield, as depicted in Fig.\,\ref{Fig_BF}. In \textbf{Scenario 1}, the fidelity of the distilled polarization state is contingent on the fidelity of the initial state in both DoFs $F^\prime_p=\frac{F_pF_a}{F_pF_a+(1-F_p)(1-F_a)}$ ($F_a$ is the initial fidelity of the auxiliary DoF, i.e., spatial-mode in PSM, energy-time in PET, frequency in our study). Notably, it demonstrates a positive gain only in the region of $F_a\in[0.5,1]$, while it does not fall below $Y=50\%$ for the range of $F_p\in[0.5,1]\,\&\, F_a\in[0.5,1]$ or $F_p\in[0,0.5]\,\&\,F_a\in[0,0.5]$. Contrary to \textbf{Scenario 2}, in cases involving PF error within polarization, it becomes entirely non-distillable, resulting in $F_p^{\prime}=F_p$. However, similar to the previous scenario, the application of a Hadamard operation to the polarization entanglement can lead to an equivalent outcome, aligning with \textbf{Scenario 1}. Upon considering all errors together in polarization, a simultaneous positive gain ($G > 0$) and reasonable yield ($Y>50\%$) can only be achieved when both $F_p>0.5$ and $F_a>0.5$, as demonstrated in Fig.\,\ref{Fig_BF}(b,d) (with $A=10\%$). Generally, there exists a trade-off between gain and yield, and it is evident that achieving high fidelity entails discarding a significant proportion of error counts.

Next, let us consider the distillation rate $\xi$, which consists of three parts: 1) the generation efficiency of the entangled source, 2) the transmission efficiency of the channel, and 3) the yield of the distillation protocol. Compared to the original (or two-copy) ED scheme, the hyperentanglement-based ED has a much higher distillation rate. For typical experimental parameters (a mean number of photon pairs per pulse of $p=0.001$ at a repetition of 76 MHz, a yield of $Y=0.8$) and two 100 km fiber-based quantum channels, the PET single-copy distillation rate outperforms the two-copy scheme by 7 orders of magnitude\,\cite{ecker_experimental_2021}. Due to the higher loss of multi-mode fiber (MCF), the PSM single-copy distillation rate is 5 orders of magnitude higher than the two-copy scheme\,\cite{hu_long-distance_2021}. In our case, assuming frequency conversion efficiency $\eta=100\%$, which is theoretically attainable, we can achieve a distillation rate similar to that of PET. However, in reality, the conversion efficiency is reduced due to coupling losses and component losses. We note that a recent experimental result has achieved a single-photon frequency conversion efficiency of approximately 50\%\,\cite{van_leent_long-distance_2020}. In this case, the yield will be slightly lower, the distillation rate is still comparable with the PET protocol. 

\section{Proposed implementation}

\begin{figure}[!t]
\centering
\includegraphics[width=3.5in]{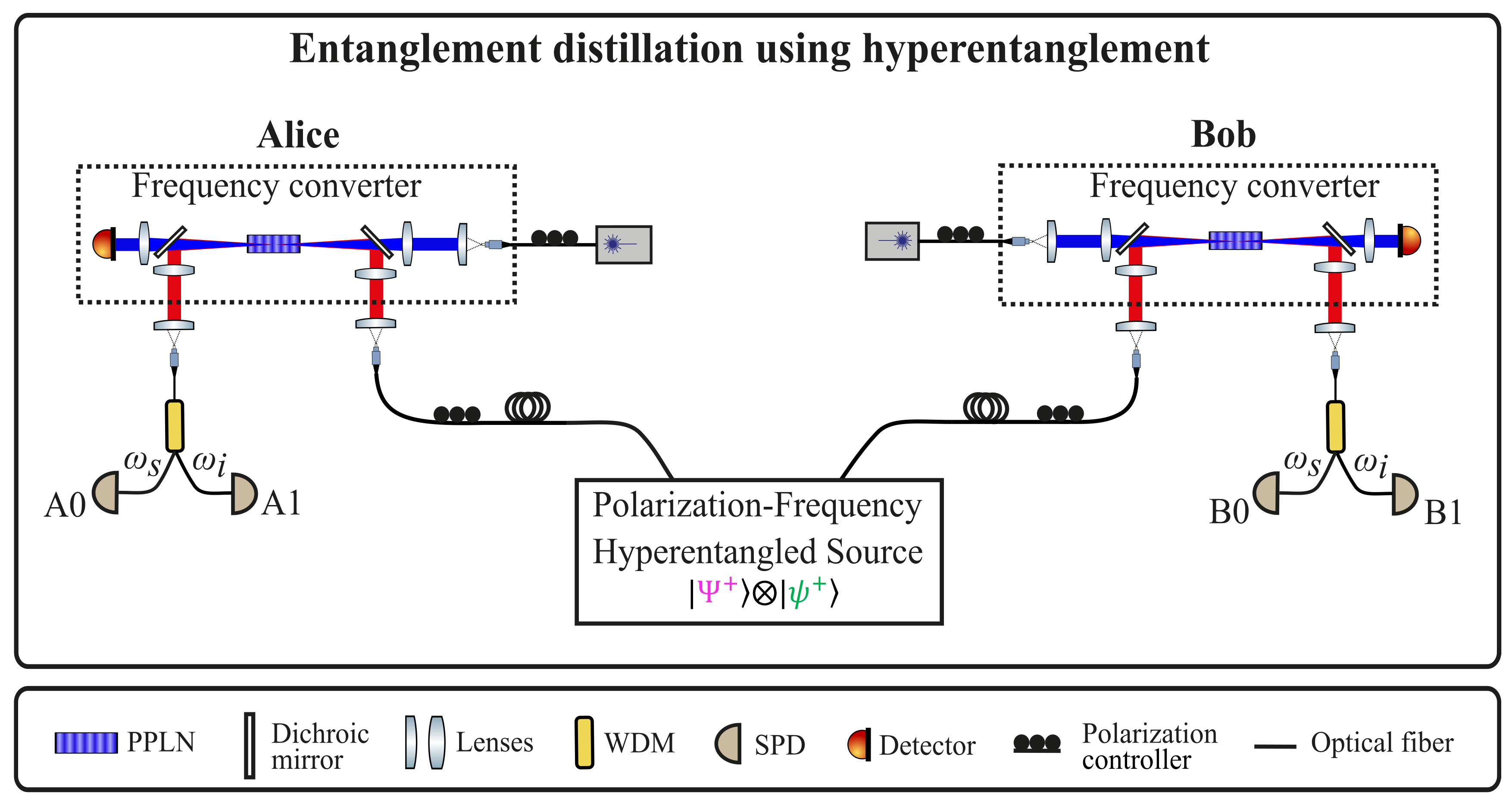}
\caption{Proposed experimental implementation for entanglement distillation in polarization and frequency. The hyperentangled state $|\Psi^+\rangle\otimes|\psi^+\rangle$ is first generated by the entanglement source where pairs of entangled photons are created in a periodically poled silica fiber (PPSF)\,\cite{chen_compensation-free_2017}. Then the entangled photons are single-mode coupled and guided to Alice and Bob. After transmission through a noisy channel, entanglement distillation is performed. The entanglement distillation operation employs polarization-dependent frequency converters as the ``CNOT" gates and converts the target qubit (frequency) according to the control qubit (polarization). The photons collected in paths A0, A1, B0, and B1 are detected and time tagged for post-selection.}
\label{Fig_exp}
\end{figure}

% description of the hyperentangled source
Hyperentanglement in polarization and frequency can be generated straightforwardly in fiber and nonlinear waveguides via nonlinear processes such as SPDC and spontaneous four-wave mixing (SFWM). As shown in Fig.\,\ref{Fig_exp}, we propose to use a periodically poled silica fiber (PPSF)\,\cite{zhu_direct_2012} to produce highly entangled ponton pairs in these two DoFs\,\cite{chen_compensation-free_2017,chen_recovering_2020,chen_telecom-band_2022}. An effective second-order nonlinearity $\chi^{(2)}$ is artificially induced along the PPSF through the process of thermal poling, and the quasi-phase-matching allows for the direct generation of polarization-entangled photon pairs in PPSF at room temperature by exploiting type-II phase-matched SPDC\,\cite{zhu_direct_2012,chen_compensation-free_2017}. A pair of orthogonally polarized photons are generated by the down-conversion of a pump photon whose polarization is along the V axis of the PPSF. The photon pairs generated by the PPSF have a continuous spectrum with broad bandwidth and are naturally endowed with frequency entanglement due to energy conservation. The hyperentangled source generates one pair of photons in the hyperentangled state $|\Psi^+\rangle_{ab}\otimes|\psi^+\rangle_{ab}$. After the photons transmit to Alice and Bob over single-mode fibers, each photon's polarization and frequency states become mixed, and the overall state is described by Eq. (4). Furthermore, we assume polarization alignment and stabilization of the quantum channel is performed as a standard practice\,\cite{ferreira_da_silva_proof--principle_2013,wei_high-speed_2020}.

% description of frequency converter
The frequency converters (FC) act as the single-photon ``CNOT" gate. To be useful in our distillation scheme, FC must (i) allow frequency translation between the signal and idler photons, (ii) preserve the polarization entanglement of the original state, and (iii) must be highly efficient while not introducing additional unwanted "noise" photons. FC using three-wave mixing (TWM) in second-order $\chi^{(2)}$ nonlinear optical media or four-wave mixing (FWM) in third-order $\chi^{(3)}$ nonlinear optical media can satisfy the above requirements\,\cite{mcguinness_quantum_2010,pelc_long-wavelength-pumped_2011,ramelow_polarization-entanglement-conserving_2012,bock_high-fidelity_2018}. For experimental implementation, we propose to realize the frequency conversion by difference frequency generation (DFG) in a periodically poled lithium niobate (PPLN) waveguide, though other nonlinear frequency conversion processes can be used as well. As shown in Fig.\,\ref{Fig_exp}, the classical pump laser is combined with the signal/idler photons at a dichroic mirror (DM). Then, they are coupled into the type-0 quasi-phase-matched (V$\to$V+V) PPLN waveguide. If the polarization state of the signal/idler photon is horizontal, then it will not meet the phase-matching condition of the PPLN, and no frequency conversion occurs. If, on the other hand, the polarization state of the signal/idler photon is vertical, the signal (or idler) photon will be frequency converted to the idler (or signal) frequency, i.e., ($|\omega_s\rangle|V\rangle\leftrightarrow|\omega_i\rangle|V\rangle$). This polarization-dependent frequency conversion process is an equivalent ``CNOT" operation between the polarization qubit and the frequency qubit. 

After the ``CNOT" operations at Alice and Bob, the photons are collected and separated by two wavelength splitters (WDMs), with signal frequency ($\omega_s$) being directed to A0 and B0, and idler frequency ($\omega_i$) being directed to A1 and B1. The successful distillation of the control qubit (polarization) is heralded by the measurement outcomes of the target qubit (frequency). In this case, either A0-B0 coincidences (i.e., both Alice and Bob detect $\omega_s$) or A1-B1 coincidences (i.e. both detect $\omega_i$) will be postselected, corresponding to frequency $|\phi^{\pm}\rangle_{ab}$ states. All other detection events will be discarded. Theoretically, a high conversion efficiency $\eta$ close to unity can be expected for a frequency converter. Experimentally, however, the highest external device efficiency reported so far is 57\% which is limited by the transmission through optical elements, fiber coupling, waveguide coupling, and spectral filtering\,\cite{van_leent_long-distance_2020}. Assuming two FCs utilized in the scheme are identical and taking the frequency conversion efficiency $\eta$ into consideration, we can obtain a distilled state with the fidelity: 
\begin{equation}
F_p^{\prime} = \begin{cases}
\frac{F_pF_f}{F_p+(1-\eta)(1-F_p)}, & (\textbf{Scenario 1})\\[8pt]
F_pF_f+(1-F_p)(1-F_f), & (\textbf{Scenario 2})\\[8pt]
\frac{F_pF_f+A(1-F_f)}{F_p+A+(1-\eta)(B+C)}. & (\textbf{Scenario 3})
\end{cases}
\label{eq9}
\end{equation}
Note that the definitions of these scenarios are provided in Section III. 

Our distillation scheme has some great practical advantages. First, our polarization-frequency hyperentanglement source using an all-fiber configuration\,\cite{chen_compensation-free_2017} could be very compact as compared to realizations using the spatial-mode DoF, which additionally requires modifications of the SPDC source\,\cite{hu_long-distance_2021}. 
Second, the long-distance distribution of polarization and spatial-mode hyperentanglement has difficulty maintaining coherence and phase stability between different paths. However, the polarization-frequency hyperentanglement is insensitive to decoherence as both $|\psi^+\rangle$ and $|\psi^-\rangle$ in frequency DoF can be used for distillation, so essentially the polarization fidelity $F^{\prime}_p$ of the distilled state is independent of frequency fidelity $F_f$ and our distillation scheme works as long as the channel remains linear. 
Third, we can use standard telecommunication fibers to transmit polarization-frequency hyperentangled photons, instead of multi-core fibers for polarization-spatial-mode hyperentanglement, so the distillation scheme we introduced can easily be integrated into the existing classical and quantum communication network. Moreover, the frequency is much more robust during transmission in in-field implementations, resulting in much higher quality entanglement in polarization DoF after entanglement distillation. 
Fourth, frequency entanglement, the same as the energy-time entanglement, arises from the energy conservation in the SPDC process. However, the scheme using energy-time DoF requires two imbalanced Mach-Zehnder interferometers and needs active phase stabilization, which increases the complexity of the whole setup. Our scheme using frequency DoF is insensitive to channel length changes, so there is no need for recalibration. 
Fifth, in terms of ultimate transmission rates, the time delay between the arrival times of the entangled photons limits the maximum bit rate that can be transmitted in polarization-energy-time hyperentanglement protocol. Polarization-frequency hyperentanglement, in contrast, has no such limitation and is therefore a promising approach for achieving high bit-rate communication.

% Advantages
% Limitations: the frequency conversion efficiency, competitive SPDC process which introduces noise (?? not decided yet)
% Noise sources which can be attributed to the QFC are spontaneous parametric down-conversion (SPDC) of the high-intensity field and Raman-scattering in the nonlinear material. The latter is suppressed by choosing the high-intensity field’s wavelength to be far off the output wavelength. Furthermore, more precisely manufactured materials can decrease the noise generated by the SPDC process in periodically poled media. 

\section{Conclusion}
In conclusion, we proposed an entanglement distillation protocol exploiting hyperentanglement in the polarization and frequency DoFs. We analyzed different error types and characterized the post-distillation states in terms of the fidelity gain, yield, and distillation rate. These theoretical results are important for future experimental implementations.
As our protocol shows, the fidelity gains are moderate in low-noise settings, while large gains are obtained in a high-noise regime where it really matters. We attribute this performance advantage to the relative robustness of the frequency DoF to common channel impairments as compared to PET and PSM schemes. This enables a significant increase in state quality in noise-dominated scenarios, recovering the potential for quantum communication applications in otherwise unattainable regimes.
Compared to the PET and PSM approaches proposed in Ref\,\cite{ecker_experimental_2021,hu_long-distance_2021}, our polarization-frequency scheme has certain significant practical advantages: a) Unlike in the case of PSM where multi-core fibers have to be used for transmitting multiple spatial modes, we can use standard telecommunication fibers to transmit polarization-frequency hyperentangled photons robustly; b) Unlike in the case of PET, our scheme does not involve interferometers and therefore no active phase stabilization is needed; c) Our scheme is insensitive to decoherence of frequency entanglement, making it a promising approach for practical applications; d) As long as the channel is linear, frequency qubits do not suffer bit flip errors during transmission, resulting in high-quality entanglement in polarization degree of freedom after one-step distillation.
% more potential applications
Improved entanglement quality after entanglement distribution in a quantum network can have a number of benefits. For example, it can improve the security of the entanglement-based quantum key distribution (QKD) protocol, or the fidelity of the teleported state, result in higher quality and robustness of quantum error correction, or improve the accuracy and efficiency of quantum simulation. Overall, our proposed polarization-frequency hyperentanglement distillation protocol has the potential to unlock many previously inaccessible benefits in quantum communication, cryptography, computing, and simulation.

\section*{Acknowledgment}

This research is supported by DEVCOM Army Research Laboratory Grant W911NF-20-2-0242 and by NSERC Alliance Grant (ALLRP 569583 EU-Canada joint project HyperSpace). 

\appendix*

\section{}
For \textbf{Scenario 3}, the PF error coexists alongside BF and BPF errors,  forming a combined error structure represented as $\rho^p_\text{err}=A|\Psi^-\rangle_{ab}\langle\Psi^-|+B|\Phi^+\rangle_{ab}\langle\Phi^+|+C|\Phi^-\rangle_{ab}\langle\Phi^-|$. Consequently, the polarization fidelity of the distilled state is expressed as $F_p^{\prime}=\frac{F_pF_f+A(1-F_f)}{F_p+A}$. The fidelity gain is depicted in Fig.\,\ref{Fig_appendix} for various values of $A$. If the BF or BPF error predominates ($A=10\%$), an analogous protocol to \textbf{Scenario 1} can be executed. Conversely, if the PF error predominates ($A=90\%$), an analogous protocol to \textbf{Scenario 2} can be executed. By applying a Hadamard gate, the fidelity gain demonstrates analogous trends showcased in Fig.\,\ref{Fig_appendix}(a) ($A=10\%$).

\setcounter{figure}{0}
\makeatletter 
\renewcommand{\thefigure}{A\@arabic\c@figure}
\makeatother
\begin{figure*}[!ht]
\centering
\includegraphics[width=\linewidth]{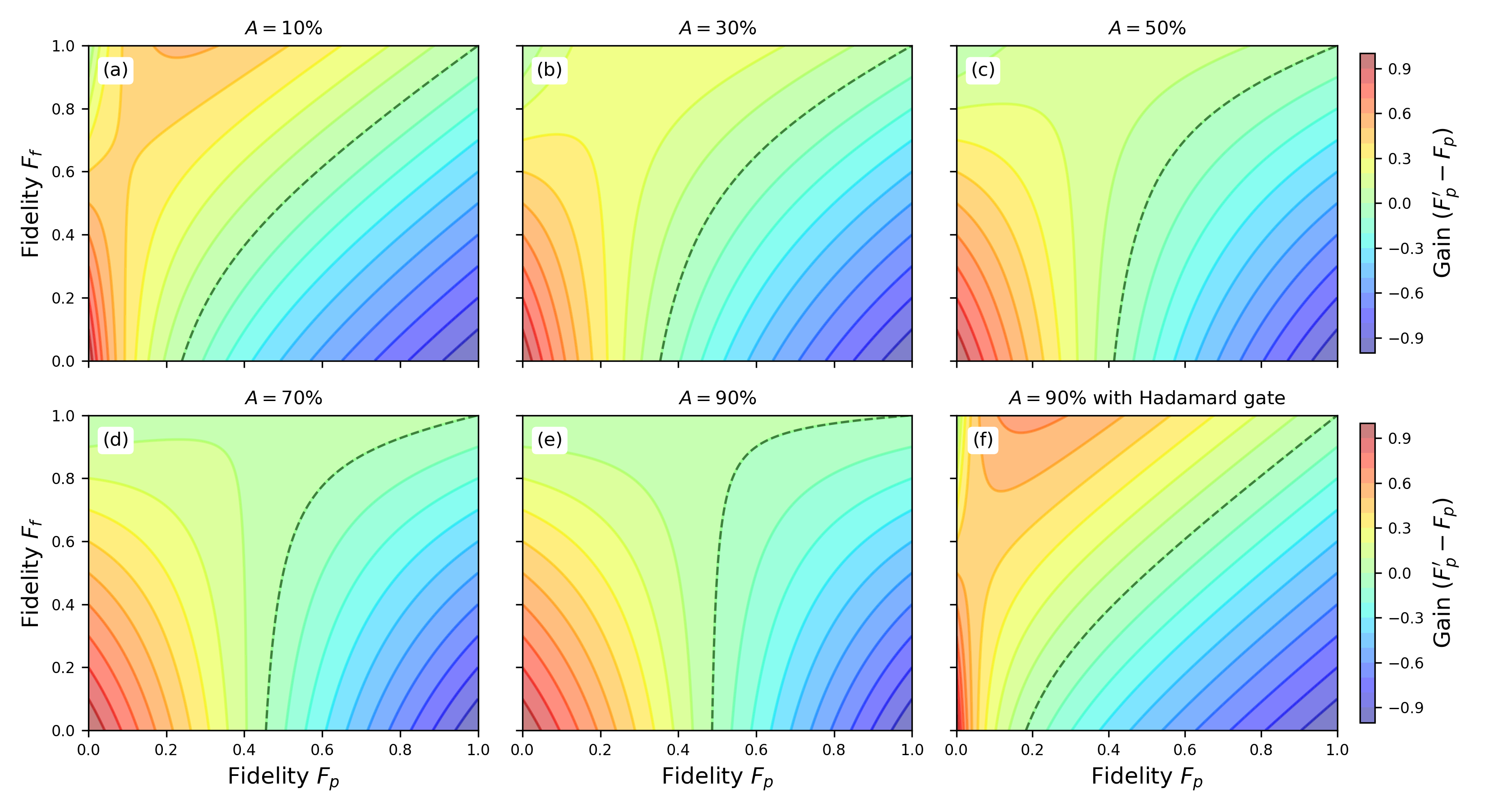}
\caption{The distilled polarization fidelity gain $G$ for \textbf{Scenario 3 (all errors together)} where we only consider PF error in the frequency qubit. We vary the value of $A=10\%, 30\%, 50\%, 70\%, 90\%$. When the PF error predominates ($A=90\%$) (e), an analogous protocol to \textbf{Scenario 2} can be executed and the fidelity gain (f) demonstrates analogous trends with (a) ($A=10\%$).}
\label{Fig_appendix}
\end{figure*}

\bibliography{ref.bib}% Produces the bibliography via BibTeX.

\end{document}